August 23, 2013

**Extended-Lorentz Quantum-Cosmology Symmetry Group**

**U(1) × SD(2,c)$_L$ × SL(2,c)$_R$**

Geoffrey F. Chew

*Theoretical Physics Group*
*Physics Division*
*Lawrence Berkeley National Laboratory*
*Berkeley, California 94720, U.S.A.*

## Summary

Unitarily representable by transformations of Milne quantum-universe (MQU) Hilbert-space vectors is a 9-parameter 'extended-Lorentz' Lie group whose algebra comprises 9 conserved MQU-*constituent* ('*quc*') attributes: electric charge, energy, spirality, 3-vector momentum and 3-vector angular momentum. Commutation with the full symmetry algebra by the 3-element Lorentz-*extending* sub-algebra identifies any *quc* by its (*permanent*) trio of charge, spirality and energy integers.

Milne's redshift-specifying 'universe age' is invariant under the MQU symmetry group. Also invariant is the (elsewhere specified) universe hamiltonian--a self-adjoint age-dependent Hilbert-space operator (*not* a symmetry-algebra member) that generates universe evolution with increasing age through a 'Schrödinger' (first-order) differential equation.

Remark: Ontological (*not* mathematical) language recognizes the 'objective reality' of a 'particle'—e.g., a photon. An MQU particle is an approximately-stable 'relationship' between *different quc*s—a 'marriage' whose persistence allows 'recognition' by *intermediate*-scale *quc* aggregates endowed with 'consciousness'. 'Intermediate scale' means *particle-scale-huge* while *Hubble-scale-tiny*.

**Introduction**

Dirac quantum theory--based on self-adjoint Hilbert-space operators--has for physics been impeded by the absence of unitary *finite*-dimensional Lorentz-group representations. The Gelfand-Naimark (GN) unitary Hilbert-space *infinite-dimensional* representation, [1] although *unusable* by a physics founded on positive-energy objective reality, applies *cosmologically* to a *Milne* (redshifting, 'big-bang') [2] *quantum universe* (MQU} where *constituent* energies may be positive *or* negative and where *total* energy vanishes. [3]

Milne's cosmological application of the Lorentz group--to a universe spacetime whose hyperbolic 3-space is *invariantly* metricized, is *profoundly* different from the physics application of the Lorentz group--to a spacetime where 3-space is Euclidean with *non-invariant* metric. Widely-misunderstood Milne cosmology associates 'redshift' so directly to universe age as to render redshift 'trivial'.

 Four-dimensional *Milne spacetime* occupies the *interior* of a *forward* lightcone, where the positive (arrowed) 'age', $\tau$, of any location is its 'Minkowski distance' from the lightcone vertex. MQU is governed by the symmetries of a here-specified 9-parameter Lie group that we designate as U(1) ×SD(2,c)$_L$ × SL(2,c)$_R$—a product of *four* individually-semisimple subgroups, three of which, each  1-parameter,  commute with the entire group.  Universe age (*not* a Hilbert-space operator) and universe hamiltonian [3] (a self-adjoint operator, although not a member of the group algebra) are both invariant under the 9-parameter MQU symmetry group.

A 2×2 unimodular-matrix meaning for the subscripts L and R in the foregoing notation will here be explained, as well as use of the symbol D in a notational location familiarly occupied by symbols such as U or L. Each of the nine members of the MQU algebra is (Stone) representable by a self-adjoint Hilbert-space operator. The (conserved) algebra corresponds to electric charge, 'spirality', energy, and a 6-vector (nonabelian) combination of momentum and angular momentum.

Particle physics is a local positive-energy restricted-scale approximation to MQU that ignores (redshifting) universe expansion. Not yet established but plausible is an *approximate* relation between charge, spirality and baryon number that accompanies approximate CP and CPT particle-physics symmetries. Reference (3) addresses both the meaning of spirality and the physics 'spatial-flattening' of MQU symmetry to the 10-parameter group that Wigner associated to the name of Poincaré.

The present paper and Reference (3) both employ the (pronounceable) acronym '*quc*' to denote an 'MQU constituent'. Each *quc* separately represents the symmetry group. The total number of *quc*s, although huge, is finite and constant--*age* independent because the hamiltonian lacks *quc* annihilation or creation operators. Each of the nine group generators represents a 'nameable' conserved *quc* attribute. The integer-valued spirality attribute distinguishes odd-integer 'fermionic' *quc*s from even-integer 'bosonic'.

Ojective reality—positive-energy temporally-stable spatially-localized 'measurement-accessible' current density--involves at least *two qucs*. A single *quc* cannot represent 'matter'—the definition of which requires some stable self-sustaining *relationship* between *different qucs*.

**Fiber-Bundle Factorization of a *Quc*'s 6-Dimensional Unimodular 2×2 Matrix Coordinate**

The 6-dimensional complex-unimodular 2×2 matrix coordinate of a *quc* (uncovered by GN in a purely-mathematical non-cosmological context [1]) admits factorization into a *product* of individually-unimodular 3-dimensional *unitary* and *positive-hermitian* 2×2 coordinate matrices. The unitary factor represents the unmetricized ('directional') fiber of a bundle whose metricized (geometric) base space is represented by the positive-hermitian factor. Either order of the two factors is possible with the *same* unitary factor, but when the hermitian factor stands on the right it is *different* from, although unitarily equivalent to, the hermitian factor when standing on the left.

Three 'Euler' angles,

$$0 \le \varphi' < 4\pi, 0 \le \vartheta < \pi, 0 \le \varphi < 2\pi, \tag{1}$$

specify, collectively, the unitary 2×2 matrix,

$$\boldsymbol{u} = exp\ (i\boldsymbol{\sigma}_3\varphi'/2)\ exp\ (i\boldsymbol{\sigma}_1\vartheta/2)\ exp\ (i\boldsymbol{\sigma}_3\varphi/2). \tag{2}$$

The matrix (2) is isomorphic to to a unit-radius 3-sphere.

In Formula (2) the symbols $\boldsymbol{\sigma}_1$ and $\boldsymbol{\sigma}_3$ denote (Pauli) hermitian traceless self-inverse *real* 2×2 matrices, each with determinant -1, $\boldsymbol{\sigma}_3$ being diagonal and $\boldsymbol{\sigma}_1$ off-diagonal. (*Any* boldface symbol in the present paper denotes a 2×2 matrix.) If the full 6-dimensional unimodular *quc*-coordinate matrix is denoted by the symbol $\boldsymbol{a}$, then

$$\boldsymbol{a} = \boldsymbol{u}\ \boldsymbol{h}_R = \boldsymbol{h}_L\boldsymbol{u}, \text{ with } \boldsymbol{h}_L= \boldsymbol{u}\ \boldsymbol{h}_R\ \boldsymbol{u}^{-1}. \tag{3}$$

A positive-trace hermitian 2×2 matrix may represent either a right or left Lorentz positive 4-vector. A right 4-vector transforms under the group SL(2,c)$_R$ by multiplication from the right by the unimodular matrix representing this group *and* multiplication from the left by the hermitian conjugate of this same matrix. A left 4-vector transforms under the group SL(2,c)$_L$ by multiplication from the left by the unimodular matrix representing the latter and multiplication from the right by this matrix's hermitian conjugate.

**Hyperbolic Base-Space Metric**

Milne's Lorentz-invariantly-metricized hyperbolic 3-space is isomorphic to the 'base space' of a (classical) fiber bundle that requires reference neither to Hilbert space nor to '*quc*'. Reference (3), on the other hand, represents the invariant MQU hamiltonian's *quc* kinetic-energy as a 'hyperbolic laplacian'—a self-adjoint Dirac operator acting on complex normed functions of *quc* base-space location.

The MQU kinetic-energy operator, a GN-uncovered positive function of extended-Lorentz-group Casimirs, [1] maintains such fiber-ignoring base-space meaning when fiber space has dimensionality 2 rather than 3. Reduction of fiber dimensionality will below be related to the meaning of 'spirality'.  The discrete meaning of both electric charge and spirality involves Hilbert space.

The combination of *quc* Hilbert space with *quc* classical fiber-bundle we call '*quc* fiber package'. This paper will identify a package with 2-dimensional fiber. The present section, however, addresses two alternative coordinations of a 6-dimensional 'classical' bundle where the fiber occupies a 3-sphere. Here, despite our use of Pauli matrices and complex numbers, we are ignoring Hilbert space.

Unitary equivalence of right and left hermitian base-space coordinate is conveniently representable in Pauli-matrix notation through 3-vector inner products.  The imaginary Pauli matrix,

$$\boldsymbol{\sigma}_2 = -i\boldsymbol{\sigma}_3\boldsymbol{\sigma}_1, \tag{4}$$

also hermitian ($\boldsymbol{\sigma}_3$ and $\boldsymbol{\sigma}_1$ anticommute), self-inverse, traceless and with determinant -1 , combines with $\boldsymbol{\sigma}_1$ and $\boldsymbol{\sigma}_3$ to define a 'handed matrix 3-vector'. Defining a unit-magnitude 'direction 3-vector' $n$ by the ordered set of components

$$n_1 = sin\Theta\, cos\phi,\ n_2 = sin\Theta\, sin\phi,\ n_3 = cos\Theta\ , \tag{5}$$

with

$$0 \leq \Theta < \pi,\ 0 \leq \phi < 2\pi, \tag{6}$$

the symbol $\boldsymbol{\sigma} \cdot n$ represents the 3-vector inner product

$$\boldsymbol{\sigma} \cdot n\ \equiv\ n_1\boldsymbol{\sigma}_1 + n_2\boldsymbol{\sigma}_2 + n_3\boldsymbol{\sigma}_3, \qquad (n \cdot n = 1). \tag{7}$$

The positive-hermitian (base-space) factor of the 6-dimensional *quc*-coordinate unimodular matrix, when standing to the right (left) of the unitary (fiber) factor $\boldsymbol{u}$, will be denoted by the symbol $\boldsymbol{h}_R$ ($\boldsymbol{h}_L$). *Any* positive-hermitian unimodular 2×2 matrix may be written in the form $exp(-\frac{1}{2}\beta\ \boldsymbol{\sigma} \cdot n)$, with $\beta \geq 0$.

We shall denote the left (right) fiber-bundle base-space (positive-hermitian) factor by the symbol: $exp(-\frac{1}{2}\beta\ \boldsymbol{\sigma}\cdot n_{L(R)})$. It follows from (3) that $\boldsymbol{\sigma}\cdot n_L = \boldsymbol{u}\ \boldsymbol{\sigma}\cdot n_R\ \boldsymbol{u^{-1}}$—the left and right directional (unit) 3-vectors being related by a rotation that leaves $\beta$ unchanged. The symbols $\beta_{R(L)}$ will sometimes be employed to denote the 3-vectors $\beta n_{R(L)}$.

Right-left invariance of (non-negative) $\beta$ reflects this symbol's interpretability as 'shortest distance' between two different *quc* locations in the 3-dimensional hyperbolic base space, *one* of the two locations being at this space's *origin*.

The hyperbolic 3-dimensional fiber-bundle base space, occupied by a (fixed) finite although huge set of *quc*s at some fixed value of age, is invariantly metricized by

$$(d\beta)^2 + sinh^2\beta\ (dn_L\cdot dn_L = dn_R \cdot dn_R), \qquad (8)$$

with

$$dn_{R(L)} \cdot dn_{R(L)} = (d\Theta_{R(L)})^2 + sin^2\ \Theta_{R(L)}\ (d\phi_{R(L)})^2\ . \qquad (9)$$

This paper will later denote simply by $n$ a unit 3-vector equal to $n_R$, but *any* coordination of 3-dimensional base-space maintains Milne's (redshift-stipulating) hyperbolic (*not* elliptic) curvature.

Any 'Lorentz boost', either from right or left, when applied to *all quc* coordinates, merely shifts base-space origin to a new location, without altering the relationship between different *quc*s (such as the spatial distance between members of a pair). More generally, *any* element of $SL(2,c)_L \times SL(2,c)_R$ , applied to *all quc* coordinates of MQU, leaves the universe unchanged.

Although the present paper considers only right 4-vectors, it attends to a 2-parameter abelian left-acting group SD $(2,c)_L$ , a subgroup of $SL(2,c)_L$ that transforms a *quc*'s coordinate matrix $\boldsymbol{a}$ by multiplication from the left by a *diagonal* 2×2 complex unimodular matrix. The reader may understand the meaning of the symbol D either as 'diagonal matrix' or as 'displacement of a complex number'. (In the following section the displaced complex *quc* coordinate will be denoted by the symbol $s$.)

The nonabelian 6-parameter group $SL(2,c)_R$ transforms the coordinate $\boldsymbol{a}$ through *right* multiplication by a unimodular complex 2×2 matrix. The groups SD $(2,c)_L$ and $SL(2,c)_R$ commute with each other. The factorization $\boldsymbol{a} = \boldsymbol{u}\ \boldsymbol{h}_R$ of the 6-dimensional *quc* coordinate matches our later Hilbert-space representation of a 9-parameter group that is isomorphic to $U(1) \times SD(2,c)_L \times SL(2,c)_R$, rather than to this group's L↔R equivalent. (In their representation of SL(2,c), GN [1] made the arbitrary choice between L and R that we employ here. The reader, if a physicist, is warned that in representing SU(2) Wigner made the *opposite* choice.)

**Alternative Factorizations of the *Quc*-Coordinate-Matrix; Spirality**

An alternative to Formula (3) is a factorization of the (6-dimensional) unimodular 2×2 complex *quc*-coordinate matrix *a* into a product of *three* unimodular 2×2 matrices, each coordinating the manifold of a 2-parameter abelian SL(2,c) subgroup (acting from *either* right or left):

$$a(s, y, z) = exp(-\boldsymbol{\sigma_3} s) \times exp(\boldsymbol{\sigma_+} y) \times exp(\boldsymbol{\sigma_-} z), \quad\quad (10)$$

where the real-matrix pair $\boldsymbol{\sigma_\pm}$ is defined as ½($\boldsymbol{\sigma_1}$± $i\boldsymbol{\sigma_2}$) and each of the symbols *s, y, z* denotes a complex variable.

The leftmost factor in (10) is a *diagonal* 2×2 matrix which may itself be written as the product, $exp(-i\boldsymbol{\sigma_3}Im\,s) \times exp(-\boldsymbol{\sigma_3}Re\,s)$ , of a commuting pair of unitary and positive-hermitian unimodular diagonal 2×2 matrices. It is useful to define a 5-parameter *quc*-coordinate matrix

$$b \equiv exp(i\boldsymbol{\sigma_3}Im\,s) \times a$$

$$, \quad\quad = exp(\boldsymbol{\sigma_3}Re\,s) \times exp(\boldsymbol{\sigma_+} y) \times exp(\boldsymbol{\sigma_-} z)\,, \quad\quad (11)$$

that depends on *Re s*, *y, z* but *not* on *Im s*.

*One* dimension of the 3-sphere unitary factor in (3) thereby becomes recognized as enjoying status *distinct* from that of a remaining dimension *pair*. [The 3-sphere of Formula (1) is the product of an 'ordinary' 2-sphere and a 4π-circumference circle.] The distinguished fiber coordinate, *Im s*, is Dirac conjugate to the self-adjoint operator representing the symmetry-group generator we call 'spirality'. [3] The (ordinary) 2-sphere 'remainder' of *quc*-fiber 3-space we call '*quc* velocity-direction space'.

**Positive Right 4-Vectors that Specify *Quc* Location in Base and Velocity-Direction Spaces**

Two right 4-vectors, one positive-timelike and one positive lightlike, are equivalent to a quintet of real *quc* coordinates (the *y, z* complex-coordinate pair plus *Re s*) that specify the 2×2 unimodular matrix *b*. In the corresponding '*quc* package' neither a *quc*'s location in 3-dimensional metricized package base space nor its location in a 2-dimensional *quc* velocity-direction space depends on *Im s*. Locations in base-space and velocity-direction space are collectively equivalent to *b*.

Formulas (3) and (11) together expose the *positive-hermitian* unimodular 2×2 matrix,

$$B \equiv b^\dagger b \quad\quad (12)$$

$$= e^{-\beta\,\boldsymbol{\sigma}\cdot n}, \quad\quad (13)$$

as a dimensionless positive-timelike right 4-vector. The dimensionful positive factor $\tau$ (age) then allows the symbol $\boldsymbol{x} \equiv \tau\boldsymbol{B}$ to denote a 4-vector which locates a *quc* within Milne spacetime—the interior of a forward lightcone—by prescribing the *quc*'s displacement from the lightcone vertex. In a more familiar notation the 4 components of $\boldsymbol{x}$ are $\tau \cos\beta$, $\tau n \sin\beta$.

Complementing dimensionless $\boldsymbol{B}$, which coordinates a fiber-bundle's metricized base space, is a second dimensionless positive right 4-vector—this one lightlike--to be denoted by the symbol $\boldsymbol{v}$ and coordinating a 2-dimensional unmetricized fiber. The 4-vector pair $\boldsymbol{B}$, $\boldsymbol{v}$ is equivalent to $\boldsymbol{b}$—an equivalence related below to invariant 4-vector inner products. (The inner product of any two *positive* 4-vectors is non-negative.)

The invariant inner product of two right 4-vectors will be denoted by the symbol •. The inner product of *any* two right 4-vectors may be shown equal to the inner product of the unitarily-equivalent *left* 4-vector pair, so *either* product is invariant under SL(2,c)$_L$× SL(2,c)$_R$ and, thereby, under the extended-Lorentz group.

The *quc velocity-direction* right 4-vector is defined to be the dimensionless zero-determinant positive-hermitian matrix

$$\boldsymbol{v} \equiv \boldsymbol{b}^\dagger (\sigma_0 - \sigma_3)\boldsymbol{b}, \tag{14}$$

where the symbol $\sigma_0$ denotes the *unit* 2×2 matrix. Equivalence of the 5-dimensional coordinate matrix $\boldsymbol{b}$ to the positive 4-vector *pair* $\boldsymbol{B}$, $\boldsymbol{v}$ accompanies the trio of inner products, $\boldsymbol{B}• \boldsymbol{v} = 1$, $\boldsymbol{B} • \boldsymbol{B} = 1$ and $\boldsymbol{v} • \boldsymbol{v} = 0$, deducible by right-transforming to a special frame where $\boldsymbol{b} = \sigma_0$. Explicit evaluation of Formula (14) via (10) and (11) reveals $\boldsymbol{v}$ *independence* of $y$—the 4-vector $\boldsymbol{v}$ being determined *entirely* by $z$ and $Re\ s$.

**Fiber-Package Unitary Representation of Extended-Lorentz Group**

An MQU ray is a sum of ('tensor') products, each with a below-discussed age-independent number of factors, of single-*quc* Hilbert-space vectors that each unitarily represents the 9-parameter extended-Lorentz group. A *regular*-basis electric-charge-*Q* Hilbert-space single-*quc* vector for Age $\tau$ is a complex differentiable function $\psi_Q^\tau(\boldsymbol{a})$ with invariant (finite) norm,

$$\int d\boldsymbol{a}\,|\,\psi_Q^\tau(\boldsymbol{a})\,|^2\ . \tag{15}$$

The invariant 6-dimensional volume element (Haar measure) $d\boldsymbol{a}$ we now express through the trio (10) of complex-variable coordinates equivalent to $\boldsymbol{a}$.

In the interest of notational simplicity we shall henceforth omit the age superscript $\tau$. (Already omitted is a label to distinguish the *quc* in question from others with the same electric charge.) Further to be ignored except in *Eq*. (21) is the integer-*Q quc*-charge subscript; U(1) transformation (Kaluza-Klein) will be seen merely to shift wave-function *phase* by an increment proportional to *Q*.

The 6-dimensional Haar measure,

$$d\boldsymbol{a} = ds\ dy\ dz, \tag{16}$$

is invariant under $\boldsymbol{a} \to \boldsymbol{a}^\Gamma \equiv \boldsymbol{a}\Gamma^{-1}$, with $\Gamma$ a 2×2 unimodular matrix representing a *right* Lorentz transformation of the coordinate $\boldsymbol{a}$. The measure (16) is also invariant under analogous left transformation. The 'volume-element' symbol $d\xi$ in (16), with $\xi$ complex, means $d\,Re\,\xi \times d\,Im\,\xi$.

The Hilbert-vector norm-defining integration (15) is, *wrt Im s*, over any continuous 2π interval of *Im s*. Below we shrink the Hilbert space so that *Re s* and *Im s* enjoy similar status in vector-norm definition. The norm (16) is then not invariant under the *full left* nonabelian group of Lorentz transformations but only under the 2-parameter abelian diagonal-matrix (D) left subgroup. Invariance under SL(2,c)$_R$ is unaffected.

A symmetry transformation specified by the 2×2 complex unimodular *right*-acting matrix $\Gamma$ is *unitarily* Hilbert-space represented by

$$\Psi(\boldsymbol{a}) \to \Psi(\boldsymbol{a}\Gamma^{-1}). \tag{17}$$

Calculation shows $\boldsymbol{a}\Gamma^{-1}$ to be equivalent to

$$z^\Gamma = (\Gamma_{22}z - \Gamma_{21})\,/(\Gamma_{11} - \Gamma_{12}z), \tag{18}$$

$$y^\Gamma = (\Gamma_{11} - \Gamma_{12}z)[(\Gamma_{11} - \Gamma_{12}z)y - \Gamma_{12}], \tag{19}$$

$$s^\Gamma = s + \ln(\Gamma_{11} - \Gamma_{12}z). \tag{20}$$

We now make explicit the single-*quc* Hilbert-space representation of U(1)×SD(2,c)$_L$×SL(2.c)$_R$, an element of which is specified by a U(1)-representing angle $\omega$, $0 \le \omega < 2\pi$, an SD(2,c)$_L$-representing complex displacement $\Delta$ and an SL(2.c)$_R$-representing 2×2 complex unimodular matrix $\Gamma$. Under an $\omega, \Delta, \Gamma$-specified extended Lorentz-group element, a regular-basis single-*quc* Hilbert-space vector transforms to

$$\Psi_Q^{\omega, \Delta, \Gamma}(s, y, z) = e^{iQ\omega}\Psi_Q(s^\Gamma + \Delta, y^\Gamma, z^\Gamma). \tag{21}$$

Under the 9-parameter group the 2-dimensional volume element *ds* within the Haar measure (16) is invariant, as also is the 4-dimensional volume element *dy dz*. Such Haar-measure factorizability dovetails with the Formula (11) factorization of *quc* coordinate space.

**Periodicity in Quc Hilbert-Vector Dependence on *Re s***

Displacement in the coordinate *Re s*, at fixed *Im s*, *y*, *z and* $\tau$, displaces what we loosely call the 'time' of an individual *quc* at fixed values of 'everything else'. *Quc* energy, as a self-adjoint Hilbert-space operator representing a well-defined member of the *left* symmetry-subgroup algebra, is canonically-conjugate in Dirac sense to what we choose (with a dimensionality-endowing factor of $\tau$) to call the 'time' of this *quc*. In the QMU hamiltonian [3] each *quc*'s gravitational potential energy is proportional to the *quc*'s energy—paralleling proportionality of *quc* electromagnetic potential energy to *quc* electric charge.

Although lightlikeness of the *quc* velocity-direction 4-vector **v** allows confusion (especially with $c = 1$) between 'temporal' and 'spatial' *quc* displacement, the group algebra unambiguously distinguishes right-invariant *quc* energy from a *quc*-momentum (3-vector) component of a right 6-vector—a component that generates (fixed-age) infinitesimal *quc* displacement in some direction through curved metricized base-space.

Because the infinitesimal-displacement direction must be specified in some *fixed* right-Lorentz frame, whereas a geodesic follows a curved path, the Reference (3) invariant self-adjoint *quc* hyperbolic laplacian—a positive Casimir function of 'geodesic-following' *second* derivatives to which a *quc*'s *kinetic* energy is proportional [1] —is *not* (as in Schrödinger's flat-space Hamiltonian) proportional to the inner product with itself of a *quc*'s 3-vector momentum.

Already noted has been the explicit confirmation by Formula (20) that (fixed-*τ*,*y*,*z*) displacements in *s* are right invariant (both real and imaginary parts). They further are invariant under the 3-parameter symmetry subgroup (with energy, spirality and electric charge as generators) that defines *quc* type, despite failure to be invariant under the full left-Lorentz subgroup. A Hilbert-space *shrinkage* specifying ray *periodicity* in dependence on *Re s* (periodicity for the hand of a '*quc* timepiece') maintains *quc* capacity to represent D-left-extended right-Lorentz transformations.

We therefore diminish each *quc*'s Hilbert space by the periodicity constraint,

$$\Psi(s, y, z) = \Psi(s + 2\pi, y, z), \tag{22}$$

with a matching redefinition of Hilbert-vector norm as integration in (15) over any (single) $2\pi$ interval of both *Im s* and *Re s*. The constraint (22) specifies *integer* eigenvalues for the self-adjoint operator *M* that is Dirac-conjugate to *Re s*.

Each of the three members of the extension subalgebra that complements the 6-member SL(2,c)$_R$ subalgebra is then Hilbert-space represented by a self-adjoint operator with (integer-specified) *discrete* spectra. The *quc* energy integer *M* is joined by the charge integer *Q* and the spirality integer *N*. (The spacing between neighboring *quc* energies is $\hbar/2\tau$—minuscule in the present universe on *any* of the scales of particle physics.)

**Conclusion**

Any *quc* is distinguished from all others by its conserved-integer trio, *Q*, *N*, *M*—the first integer, *Q*, specifying *quc* electric charge, the second, *N*, specifying *quc* spirality (distinguishing fermionic *quc*s from bosonic) while the third, *M*, specifies *quc* energy (in units of $\hbar/2\tau$). The finite total number of different *quc*s—the population of the set of allowed integer trios—is addressed in Reference (3). Because, within this set, any positive integer is accompanied by the corresponding negative integer, the total value *vanishes* of universe electric charge, spirality and energy.

Also vanishing is total-universe angular momentum—QMU's version of Mach's principle. Because *quc* momentum is continuous, Hilbert-space meaning for vanishing total-universe-momentum is more subtle; the meaning is addressed in an appendix to Ref. (3).

A 'type-basis' Hilbert-space vector $\Phi_{Q,N,M}(y, z)$, for the *quc* identified by the integer trio *Q*, *N*, *M*, is a differentiable normed complex function of two complex variables. In terms of the complex displacement (linear in $\Delta$ although *z*-dependent via $\Gamma$) that is specified by the formula

$$\delta\,(\varDelta,\,\boldsymbol{\varGamma},\,z\,)\equiv\,\varDelta + ln\,(\varGamma_{11} - \varGamma_{12}z),\tag{23}$$

any *quc*-type-basis Hilbert-space vector represents the group element specified by $\omega$, $\varDelta$, $\boldsymbol{\varGamma}$ through the transformation

$$\varPhi_{Q,N,M}\,(y,z)\rightarrow e^{i[Q\omega\,+N Im\,\delta\,+\,M\,Re\,\delta\,]}\,\varPhi_{Q,N,M}\,(y^{\varGamma},z^{\varGamma}).\tag{24}$$

The role of the *quc* coordinate *Re s*, absent from (24), deserves attention in our conclusion. This coordinate is resurrectable by defining the Hilbert vector

$$\varPsi_{Q,N,M}\,(\boldsymbol{b})\equiv e^{\,-iM\,Re\,s}\,\varPhi_{Q,N,M}\,(y,z),\tag{25}$$

that recalls equivalence of the matrix coordinate $\boldsymbol{b}$ to the quintet *Re s*, *y*, *z* which coordinates the fiber bundle of 3-dimensional metricized hyperbolic base space and 2-dimensional velocity-direction fiber.

Established has been (classical) $\boldsymbol{b}$ equivalence to the right positive 4-vector pair $\boldsymbol{B}$ (base-space location) and $\boldsymbol{v}$ (velocity direction). Although both these 4-vectors depend on *Re s* and *z*, because $\boldsymbol{v}$ *fails* to depend on *y*, the latter may described *classically* as a '2-dimensional base-space *quc*-location coordinate'. Contrastingly, because *Re s* and *z* collaborate in 'fiber-package' roles that involve Hilbert space, this coordinate trio fails to admit a classical name.

Ontologically-justifiable names for 8 of 9 members of the extended-Lorentz algebra are unproblematic. The name 'spirality', assigned here to one algebra member, may or may not survive the test of usefulness.

 **Acknowledgements**

Decades of discussions with Henry Stapp and Jerry Finkelstein have led to the quantum cosmology here proposed. Also contributing have been remarks by Korkut Bardakci, David Finkelstein, Eyvind Wichmann, Bruno Zumino, and Nicolai Reshetikhin.